# Subcycle videography of lightwave-driven Landau-Zener-Majorana transitions in graphene

V. Eggers[1], G. Inzani[1], M. Meierhofer[1], L. Münster[1], J. Helml[1], R. Wallauer[2], S. Zajusch[2], S. Ito[2],

L. Machtl[1], H. Yin[3], C. Kumpf[3], F. C. Bocquet[3], C. Bao[1], J. Güdde[2],

F. S. Tautz[3], R. Huber[1], and U. Höfer[1,2]

[1] *Department of Physics and Regensburg Center for Ultrafast Nanoscopy (RUN),*

*University of Regensburg, 93040 Regensburg, Germany*

[2] *Department of Physics, Philipps-Universität Marburg, 35037 Marburg, Germany*

[3] *Peter Grünberg Institut (PGI-3), Forschungszentrum Jülich, 52425 Jülich, Germany*

**Strong light fields have unlocked previously unthinkable possibilities to tailor coherent electron trajectories, engineer band structures and shape emergent phases of matter all-optically[1,2,3,4,5,6,7,8,9]. Unravelling the underlying quantum mechanisms requires a visualisation of the lightwave-driven electron motion[10,11] directly in the band structure. While photoelectron momentum microscopy has imaged optically excited electrons averaged over many cycles of light[12,13,14,15,16,17,18], actual subcycle band-structure videography has been limited to small electron momenta[8,19]. Yet lightwave-driven elementary processes in quantum materials often occur throughout momentum space. Here, we introduce attosecond-precision, subcycle band-structure videography covering the entire first Brillouin zone (BZ) and visualize one of the most fundamental but notoriously elusive strong-field processes: non-adiabatic Landau-Zener-Majorana (LZM) tunnelling. The interplay of field-driven acceleration within the Dirac-like band structure of graphene and periodic LZM interband tunnelling manifest in a coherent displacement and distortion of the momentum distribution at the BZ edge. The extremely non-thermal electron distributions also allow us to disentangle competing scattering processes and assess their impact on coherent electronic control through electron redistribution and thermalization. Our panoramic view of strong-field-driven electron motion in quantum materials lays the foundation for a microscopic understanding of some of the most discussed light-driven phenomena in condensed matter physics.**

According to the adiabatic theorem, a time-dependent two-level quantum system remains in its ground state during a slow evolution and its eigenstates never cross[20]. In contrast, driving the system faster than its intrinsic timescale through an avoided crossing of its energy levels can create a finite occupation of the excited state via non-adiabatic transitions called LZM tunnelling[21,22,23]. In solids, this mechanism plays a key role in strong-field light-matter interaction[7,24,25,26]. Yet a direct observation of this foundational quantum effect in momentum space is still missing. Graphene's unique optical, electronic, and chemical properties[27] represent the ideal testbed for observing LZM transitions. Its quasi-relativistic electron dispersion and the robustness under irradiation with intense light fields have made this material a prime test ground for lightwave electronics[10,11], including the demonstration of first lightwave electronic logic gates[7] and the observation of Floquet-Bloch states[17,18]. This has opened revolutionary prospects for next-generation quantum control, quantum sensing, and quantum computation applications[10,11,28].

To understand and shape the intriguing opportunities offered by lightwave control of electrons, these breakthroughs call for the direct observation of how LZM transitions dictate strong-field light-matter interaction. Attosecond spectroscopy has reached sufficient temporal resolution to access these dynamics[29,30], while time- and angle-resolved photoemission spectroscopy (trARPES) in combination with momentum microscopy[31] has excelled in visualizing electron dynamics directly in the band structure of quantum materials, or in the orbitals of molecular layers[12,13,14,15,16,17,18,32]. The recent development of trARPES with subcycle temporal resolution[8,19] has demonstrated the potential of combining these approaches, but only offered a limited one-dimensional access to momentum space. To track how coherent field-driven phenomena and scattering-induced momentum redistribution conspire, a wholistic observation of carrier dynamics in the entire momentum space is imperative.

Here we explore this crucial light-matter interaction regime by expanding, for the first time, subcycle band-structure videography to the entire first BZ. The new platform overcomes previous restrictions of both, momentum microscopy and subcycle ARPES, increasing the momentum area accessible with subcycle resolution by more than two orders of magnitude, and the available field strengths by one order of magnitude. Exploiting few-cycle mid-infrared (MIR) field transients with large peak amplitudes, we drive coherent electron dynamics throughout the entire band structure of graphene.

Our novel approach allows us to reveal the simultaneous displacement and distortion of the light-driven electron distribution, uncovering the complex interplay between coherent intra- and interband dynamics underlying LZM transitions in graphene. Moreover, the possibility to prepare and track extremely non-thermal subcycle distributions in two-dimensional momentum space enables us to reveal the effect different scattering processes take on lightwave electronic dynamics.

**Subcycle lightwave currents**

When a low-intensity short-wavelength light pulse interacts with electrons in a semiconducting or insulating solid, it can perturbatively induce interband quantum leaps in momentum regions where the energy gap between the occupied valence and the unoccupied conduction bands matches an integer number of photons (Fig. 1a). In graphene, the non-trivial pseudospin texture directly influences the strength of the light-induced coupling between the lower and upper parts of the Dirac cone, leading to a distribution of the excited carriers only along the direction orthogonal to the light field[33,34].

Conversely, a sufficiently strong low-frequency electric field transient holds the potential to adiabatically accelerate charge carriers within their bands, strongly deflecting the Fermi surface in momentum space according to Bloch's acceleration theorem (Fig. 1b). During their excursion through the band structure, electrons experience a time-changing energy gap which depends on the bands' dispersive landscape. In the specific case of Dirac fermions in graphene, close to the K point, the band gap varies linearly, resulting in an avoided crossing. Therefore, the field-driven acceleration of carriers may lead to prototypical LZM transitions, leaving a markedly different occupation in momentum space compared to Fig. 1a. Furthermore, if the perturbation is periodic, the phase accumulated during the coherent evolution between subsequent non-adiabatic events – referred to as the Stückelberg phase – will result in a dynamical interference[35]. The interplay of these phenomena is called Landau-Zener-Stückelberg-Majorana (LZSM) interferometry[36].

To access this regime, we drive electrons in an epitaxially-grown monolayer of graphene on a SiC substrate[37] with intense phase-stable MIR pulses at a central frequency of 30 THz. After a delay time, $t$, ultrashort XUV pulses (central photon energy, 21.7 eV) photoemit the electrons with subcycle resolution. Their energies and momenta are detected with a time-of-flight photoemission momentum microscope (Fig. 1c, Methods). In equilibrium, the acquired photoemission signal maps the occupied band structure of graphene. Owing to the high XUV photon energy and the large acceptance angle of the momentum microscope, the entire first BZ can be imaged in a single measurement, with sub-10-femtosecond temporal resolution. Importantly, we can capture the six K points of graphene at the edge of the first BZ, where the Dirac cones are located.

Figure 2a shows several curvature-filtered (see Methods) momentum distribution maps in the vicinity of a K point at selected discrete energies. The in-plane component of the *p*-polarized XUV radiation driving the photoemission process is set parallel to the $k_x$ axis. Due to matrix element effects in the photoemission process[38], the intensity exhibits the characteristic horseshoe-like distribution for energies far below the Dirac point. The influence of the dark corridor at $k_y = 0$ and negative $k_x$ values is suppressed for energies close to the Dirac point, $E_D$, and the photoemission intensity forms a disk. To ensure that the relevant dynamics occur along a direction where the two branches of the Dirac cone have equal photoemission intensities, we apply an *s*-polarized MIR field along the $k_y$ axis. For large negative delays ($t = -70$ fs), a slice of the curvature-filtered photoelectron intensity for $k_x - K = 0$ (Fig. 2b) displays two linear bands with a slope of $\hbar v_F$ (where $v_F = 10.7$ Å/fs is the Fermi velocity) and a Fermi level located 200 meV above the Dirac point. With increasing *t*, we observe an apparent shift of the overall band structure owing to momentum streaking[19] of photoemitted electrons in the incident MIR field in vacuum. This effect can hence be employed to directly retrieve the MIR waveform at the sample surface (Extended Data Fig. 2, Methods). By compensating for this shift, we can single out the subcycle field-driven electron dynamics occurring inside the electronic band structure of graphene.

A femtosecond snapshot of the occupied band structure for $t = -13.6$ fs shows a highly non-thermal lightwave-driven electron distribution (Fig. 2b). In the upper half of the Dirac cone ($E - E_D \geq 0$), the $k_y > 0$ branch is occupied up to ~ 1 eV, while the $k_y < 0$ branch is completely unoccupied. Half a cycle of the MIR carrier wave later ($t = 6.3$ fs), the field direction is inverted, and the electron distribution is mirrored. This unequivocal hallmark of lightwave-driven acceleration in direct band structure videography is remarkable given the expected ultrashort scattering times in graphene[32,34,39,40,41,42,43,44]. The instantaneous asymmetric electron distribution in momentum space can be associated with the excitation of a microscopic current. By tracking the imbalance in the band structure occupation for positive and negative $k_y$ momenta (Fig. 2c, inset, see Methods), we extract the lightwave-driven current density and directly compare it to the driving MIR waveform (Fig. 2c). Both the amplitude envelope and the frequency of the oscillations of the current density have the same features as in the MIR transient. However, the data show a non-trivial phase shift between the two waveforms.

From Bloch's acceleration theorem[19,45], one would expect that under idealized conditions, where decoherence has not taken effect yet, the momentum accumulated by electrons follows the integral over the MIR field, resulting in a phase shift of $\pi/2$ between the current and the field. However, a Fourier analysis (see Methods) allows retrieving this phase delay with attosecond precision. It reveals an average value of $4.9 \pm 0.2$ fs, significantly smaller than the value associated with a fully coherent evolution (8.3 fs). Simulations based on the semiconductor Bloch equations (SBEs)[6] in the relaxation time approximation (Extended Data Fig. 3, Methods) reproduce this phase shift, which is associated with a mean scattering time of $9.0 \pm 0.7$ fs. As non-adiabatic LZM transitions rely on the coherent and unperturbed field-driven evolution of charge carriers, this span ultimately sets the time and length scales relevant for lightwave electronics. Despite the short scattering time, electrons in graphene can travel almost 100 Å due to their large velocity of $v_F = 10.7$ Å/fs.

**Interband coupling in electron dynamics**

Most remarkably, the current is accompanied by an increase of the total population of the conduction band (Fig. 2c, grey shaded area). Since in the doped sample interband absorption is prohibited by Pauli blocking at equilibrium, this indicates an interplay between strong-field driven intraband and interband dynamics. In the following, we unveil the microscopic details of the mechanisms at play by imaging the full 2D momentum space. With this capability, not available in previous subcycle photoemission experiments[8,19], we can directly target the coherent light-matter interaction regime and the subsequent relaxation dynamics.

Before the interaction with the MIR transient, the integrated equilibrium carrier distribution in the conduction band ($E - E_D \geq 0$) of n-doped graphene can be approximated by a disc centred about each K point (Fig. 2a). If the electrons were only subject to an accelerating electric field, all electrons would be collectively shifted along the $k_y$ direction according to Bloch's acceleration theorem[45] and we would expect to observe merely a time-dependent centre-of-mass (COM) displacement of this disc. While the measured 2D momentum maps (Fig. 3) indeed show pronounced oscillations of the COM displacement with the frequency of the driving field and a maximum amplitude of 0.078 Å$^{-1}$ (Fig. 3a,b), they also clearly reveal an additional drastic change of their shape (Fig. 3a,c). In fact, the observed deviations from the initial shape of the distribution are of the same order as the COM displacement.

Already after the first cycles of the MIR field (Fig. 3a, $t = 3$ fs) we find a surprisingly strong elongation of the Fermi surface, from 0.054 Å$^{-1}$ to 0.096 Å$^{-1}$, in the direction parallel to the field, whereas the change in the orthogonal direction is only 0.014 Å$^{-1}$ (Fig. 3a, Gaussian fits). The Fermi surface is no longer circular but takes on an elliptical shape. The time-dependent change in its $k_y$ axis (Fig. 3c, green line) reveals a direct correlation between the field strength and the degree of broadening. Additionally, the width of the electron distribution oscillates at twice the frequency of the field while the COM displacement tracks the electric field in a similar fashion as the current plotted in Fig. 2c. Because the maximum broadening coincides with the peak displacement of the COM, we can rigorously rule out motion blur as the origin of the elongation in the $k_y$ direction. Remarkably, the electron distribution encompasses the Dirac point at all delay times, even for the largest displacements.

All these intriguing dynamics can be microscopically explained by a full quantum-mechanical simulation of electronic intraband acceleration and LZM-type interband coupling on equal footing, using the semiconductor Bloch equations (SBEs)[6] (see Methods). A systematic switch-off analysis pinpoints how specific aspects of coherent strong-field light-matter interaction manifest in the observed dynamics (Extended Data Fig. 4a). The periodic COM shift of the electron distribution is clearly associated with field-driven intraband motion of charge carriers. Yet, the momentum distribution retains its circular shape and separates from the K point for large displacement if interband dynamics are switched off. Once interband coupling is included, the theory shows the same elongation of the carrier distribution as the experiment (Extended Data Fig. 4b). The theory also qualitatively reproduces the rise in carrier density in the conduction band (Extended Data Fig. 5). This confirms that interband transitions at the K point act like a well that feeds electrons into the upper part of the Dirac cone on subcycle time scales.

The observed interband transitions, however, are not described by conventional perturbative photoexcitation but are clearly a result of the strong-field nature of interband coupling under our experimental conditions. Owing to pseudospin selection rules[33,34], photoexcitation would occur in momentum regions with $|k_x - K| > 0$, located perpendicular to the polarization of the incident light (Extended Data Fig. 6a). In contrast, the region where LZM-type tunnelling can occur is determined by the field strength and the energy gap. The latter increases linearly with $|k_x - K|$ and confines the transition region to momenta $|k_x - K| \leq 0.078$ Å$^{-1}$ for a maximum field strength of 1.2 MV/cm. This momentum region is only slightly larger than the initially occupied one (Extended Data Fig. 6b-d). The coupled intra- and interband motion of the LZM scenario thus leads to the observed unidirectional elongation of the electron distribution parallel to the electric field. It also explains the modulations of the width with twice the driving frequency observed near zero delay time (Fig. 3c). Since the region close to the K point is the one with the lowest tunnelling barrier, electron injection into the upper part of the Dirac cone becomes most efficient in the few-fs time window between the maxima of the electric field and the maximum displacements of the already existing population, which is two times per optical cycle. Scattering shortens this time window, which shifts the optimal injection time (see Methods), and an increased dephasing rate in the simulations, which destroys the quantum memory between successive LZM transitions, can dampen the oscillations in the width of the distribution (Extended Data Fig. 7). Their presence in the data therefore suggests interference of selected electrons due to their accumulated phase between individual tunnelling events, emphasizing the role of quantum coherence in light-field-driven devices.

**Ultrafast momentum-resolved relaxation**

While LZM dynamics dominate at early delay times, the subsequent emergence of scattering sets the limiting time frame for fully coherent light-matter interaction. At later delay times (Fig. 3a, $t = 26.3$ fs), the initially anisotropic electron distribution broadens also along the $k_x$ direction and gradually approaches a circular shape with a radius of $\sim 0.065$ Å$^{-1}$, before it isotropically contracts again towards equilibrium ($t = 233$ fs). This temporal evolution, evident from the time-dependent changes in the two principal axes of the elliptical distribution (Fig. 3c), reflects the interplay between field-driven acceleration, LZM transitions, scattering-induced angular thermalization, and carrier relaxation. Our simulations based on the SBEs (Extended Data Fig. 8, Methods) predict distinctive momentum fingerprints for electron-electron (ee) and electron-optical phonon (op) scattering processes. 2D mapping thus allows us to identify their signatures (Fig. 4a). Neglecting all forms of scattering results in a broadening (up to 0.053 Å$^{-1}$) and displacement (~ 0.107 Å$^{-1}$) of the carrier distribution along the direction of the driving field, and the distribution remains elliptical at the end of the MIR pulse ($t =$ 233 fs), retaining the momentum hallmark of LZM transitions. In addition, the phase delay of the microscopic current (Fig. 4b) is approximately $\pi/2$, consistent with our expectations for undamped intraband motion.

Introducing electron-electron scattering, the initially strongly broadened and displaced momentum distribution ($t = 27$ fs) partially thermalizes on longer timescales ($t = 233$ fs) and shrinks to a circular shape with a radius of ~ 0.038 Å$^{-1}$. Still, the peculiar angular dependence of electron-electron scattering[40,46] retains a large ellipticity on shorter timescales while, surprisingly, the phase delay is only marginally affected. This effect may be attributed to the reduced phase space for scattering imposed by the pseudospin texture of graphene and the nature of LZM tunnelling, which results in a unidirectional excursion of the carrier distribution.

Eventually, also including scattering with Γ point longitudinal (LO) and transverse (TO) optical phonons enables the simulations to fully reproduce the experimentally observed dynamics. Already at early stages of the interaction with the pump pulse ($t = 27$ fs), the electron distribution shows a significant broadening both in the direction parallel (~ 0.045 Å$^{-1}$) and perpendicular (~ 0.037 Å$^{-1}$) to the pump field, associated with an isotropic redistribution of charge carriers. This leads to an almost circular distribution whose centre of mass is shifted by only 0.008 Å$^{-1}$ at $t = 27$ fs, significantly less than in the previous cases where it was 0.081 Å$^{-1}$. This can be attributed to a remarkable change in the phase delay (Fig. 4b), approaching a value of $\pi/4$. After the pulse ($t = 233$ fs), the distribution relaxes to a fully circular shape with a radius of ~ 0.031 Å$^{-1}$. In this situation, both the phase delay and the time-dependent broadening of the electron distribution along the $k_x$ and $k_y$ directions (Fig. 4c) are in qualitative agreement with the experiment. This observation allows us, for the first time, to identify the key limiting factor for coherent control dictated by the fastest momentum redistribution channel, which in graphene is electron–optical phonon scattering (Fig. 4b, solid blue line).

**Conclusion**

Our experiment opens the momentum horizon of subcycle photoemission over the entire first BZ of basically any quantum material. This universal tool enabled us to directly observe LZM transitions in the band structure of a solid, for the first time. Attosecond-precision videography of graphene unravels how interband LZM tunnelling couples with field-driven intraband acceleration of quasi-relativistic fermions along the Dirac cone. Simultaneously, the motion pictures visualize how specific scattering events influence and limit the fully coherent evolution of the system. Following the same principle, our comprehensive subcycle band-structure videography may soon resolve a plethora of other key open questions of modern condensed matter physics and upcoming ultrafast information technologies. Examples range from optical engineering of novel band structures and phase transitions to light field-driven sculpting of molecular orbitals.

**Methods**

**Experimental setup**

The subcycle momentum microscopy setup (Extended Data Fig. 1a) is based on two ytterbium-doped potassium gadolinium tungstate (Yb:KGW) amplifiers (Light Conversion Inc., CARBIDE). Each amplifier generates near-infrared light pulses with a duration of 250 fs, a centre wavelength of 1030 nm, and a pulse energy of 1.5 mJ, at a repetition rate of 50 kHz. In the probe arm, the second harmonic of the output of a four-stage noncollinear optical-parametric amplifier (NOPA, Light Conversion Inc., ORPHEUS OPCPA)[47] yields CEP-stabilized pulses centred at 400 nm with a pulse duration of 20 fs. These pulses are focused into a high-pressure argon gas jet to generate high-order harmonics (Extended Data Fig. 1b) in a vacuum beamline[48]. The 7th harmonic at a photon energy of 21.7 eV is spectrally selected by two multilayer mirrors (grey shaded area, Extended Data Fig. 1b). They feature a bandwidth of 0.9 nm around 57 nm and thus support a Fourier-limit of less than 10 fs. The XUV pulses enter the experimental ultrahigh-vacuum (UHV) chamber, the momentum microscope[31], through a 200-nm-thin aluminium foil, which also filters out the UV driving pulses.

Since in a time-of-flight photoemission momentum microscope the electron detection rate is limited to only about one electron per laser shot, a high repetition rate is key for keeping measurement times at a reasonable level while ensuring good statistics. In this experiment, this requirement entails the challenge of pushing the generation of the atomically strong, few-cycle phase-stable MIR pump pulses to new limits for a tabletop setup. Here, the generated phase-stable mid-infrared pulses feature, to the best of our knowledge, record-high average powers of up to 1 W and electrical peak field strengths of more than 200 MV/cm. These field transients are obtained by difference frequency generation (DFG) in a gallium selenide crystal between the outputs of two parallel two-stage (N)OPAs (Light Conversion Inc., ORPHEUS MIR)[49]. They feature pulse durations down to sub-50 fs and can be tuned between 4 μm and 15 μm. A small portion of the laser fundamental is picked off to pump a home-built single-stage NOPA. It is optimized to generate 10-fs pulses centred around 825 nm, which are utilized for electro-optic sampling (EOS) of the MIR waveforms. Extended Data Fig. 1c shows a typical MIR transient detected by EOS. A pair of wire-grid polarizers allows us to precisely control the polarization direction and power of the MIR pulses, which are then coupled into the UHV chamber through a diamond window.

The intense MIR pump (*s*-polarized) and the XUV probe (*p*-polarized) are spatially overlapped in the vacuum chamber with a custom-built off-axis parabolic mirror featuring an aperture for the transmission of the XUV pulses and are focussed onto monolayer graphene with an angle of incidence of 70°. A piezo delay stage in the excitation beam path allows us to temporally delay the MIR and XUV pulses with respect to each other while retaining an attosecond temporal precision of 230 as. The emitted photoelectrons are detected by a time-of-flight momentum microscope (Lens system: GST; Detector: RoentDek, Hex40)[50]. Multi-hit detection ensures linearity of the signal up to a repetition rate of 20 kHz. The combination of the 21.7-eV probe pulses with the large acceptance angle of the momentum microscope, allows access to momentum regions up to $\pm 2.14$ Å$^{-1}$, as compared to the $\pm 0.16$ Å$^{-1}$ in previous subcycle ARPES experiments[8,19]. The energy resolution is limited by the broad spectrum of the probe pulse. The bandwidth of the XUV pulse is tuneable, setting a limit for the energy resolution of 180 meV. The momentum resolution is better than 0.01 Å$^{-1}$. The custom-designed UHV chamber is shielded by a mu-metal liner and maintains a base pressure of $2 \times 10^{-10}$ mbar. The samples were measured at room temperature.

Inducing field-driven dynamics in the monolayer requires a field that is parallel to the plane of the monolayer. To maximize the corresponding field component the MIR pulse is *s*-polarized. The polarization of the XUV pulse is set to be perpendicular to the MIR pulse to prevent crosstalk between their fields that could introduce coherent artifacts effects in the observed dynamics. Finally, the rotation of the sample is chosen such that the MIR-induced momentum streaking in vacuum is minimized at one of the K points. This requires one K point to be at $k_\parallel = 0$, with $k_\parallel$ being the momentum direction parallel to the electric field of the MIR pulse. This alignment also rules out matrix element-induced asymmetries along $k_\parallel = 0$ in the observed momentum range.

**Graphene samples**

For sample preparation, an n-doped 6H-SiC(0001) wafer was used as a substrate for growing an epitaxial monolayer graphene with 0° rotational alignment relative to the substrate lattice (0°-EMLG). The wafer was initially degassed in UHV at 880 °C for 30 min. The temperature was subsequently increased to 1050 °C and maintained for 30 min under an external flux of Si atoms, resulting in a Si-rich (√3×√3)R30° surface reconstruction. The sample was then further annealed at 880 °C for additional 30 min, again under Si flux, resulting in the Si-rich (3×3) surface reconstruction. With this procedure the oxide layer is removed from the SiC surface and atomic cleanness is achieved. In the next step, 0°-EMLG was obtained by annealing the sample at 1400 °C for 30 min in borazine ($B_3N_3H_6$) atmosphere with a partial pressure of $1.5 \times 10^{-6}$ mbar. The borazine molecules act as surfactants on the heated SiC surface. They inhibit the formation of the conventional 30°-rotated graphene layer but instead force the homogeneous graphene layer in an unconventional 0° orientation with respect to the substrate lattice. For more details on the preparation of the 0°-EMLG, see refs. 37, 51, and 52. Following growth, the surface morphology and crystalline structure were characterized using low-energy electron microscopy (LEEM), and low-energy electron diffraction (LEED). The average width of terraces on the SiC surface is approximately 160 nm, and the surface is fully covered by 0°-EMLG. Subsequently, the sample was shortly exposed to air, transported in static vacuum (0.1 mbar) to the experimental chambers, and degassed at 350 °C for 3 h prior to the measurements. The photoemission experiments reveal that the 0°-EMLG sample is n-doped, with the Fermi level approximately 200 meV above the Dirac point.

**Electric field waveform reconstruction and streaking compensation**

The acceleration of photoemitted electrons in vacuum allows for an in-situ reconstruction of electric-field waveforms[19]. The s-polarized field causes a momentum change parallel to the surface of the sample, $\Delta k_\parallel$, which is proportional to the vector potential of the field. The electric field waveform can be calculated by:

$$\boldsymbol{E}_\parallel(t) = \frac{\hbar}{e}\frac{d}{dt}\Delta k_\parallel(t).$$

To extract $\Delta k_\parallel$, the extracted photoemission data is shifted stepwise along $k_\parallel$ and compared to a reference spectrum without a field. The absolute difference between the investigated data and the reference is calculated for every pixel and summed for each iteration. The iteration with the smallest overall error gives the value for $\Delta k_\parallel$. Repeating this procedure for every delay point, $t$, allows a reconstruction of the applied field $\boldsymbol{E}_\parallel(t)$. To refine the retrieved waveform, a second method for extracting $\Delta k_\parallel(t)$ is used. Tracking the change in the number of electrons, integrated over all detected energies (+1 eV to –7 eV), with positive or negative momentum along $k_\parallel$, yields a unit-less quantity directly proportional to $\Delta k_\parallel(t)$, which is not subject to discretization caused by the binning of the data along $k_\parallel$. The waveform that can be extracted in this way is then scaled by the peak electric field from the first method. The resulting waveform is shown in Fig. 2c.

The extracted $\Delta k_{\|}$ is also used to compensate for the streaking in the photoemission spectra. In addition to the trivial shift along $k_{\|}$, a shift in energy, $\Delta E$, is connected to momentum streaking. The energy of every electron in vacuum is given by:

$$E(\boldsymbol{k}) = \frac{\hbar^2}{2m}\boldsymbol{k}^2.$$

The change in energy caused by $\Delta k_{\|}$ can therefore be calculated with:

$$\Delta E(k_{\|}) = E(\boldsymbol{k} + \Delta \boldsymbol{k}) - E(\boldsymbol{k}) = \frac{\hbar^2}{m}k_{\|}\Delta k_{\|} - \frac{\hbar^2}{2m}\Delta k_{\|}^2.$$

Extended Data Fig. 2 displays band structure snapshots taken without an applied field (left panel) and with a strong field applied before streaking compensation (middle panel) and after streaking compensation (right panel). The streaking-induced changes of energy and momentum were compensated for in all photoemission spectra before further analysis of the occupation of the band structure.

**Curvature filtering**

The data in Fig. 2 is first processed with a 1D simple moving average (SMA) along the energy and momentum axis with window sizes of 0.59 eV and 0.08 Å$^{-1}$, respectively, for four times. The filtered data is further processed by a 2D curvature filter described by equation (11) of ref. 53 with $C_0 = 0.2$. Finally, a 2D Median filter with the window size of 0.34 eV × 0.045 Å$^{-1}$ is applied three times to clean up numerical noise. The raw data is shown in Extended Data Fig. 9.

**Current extraction and Fourier analysis of the phase delay**

The microscopic current density for each time instant, $j(t)$, is directly extracted from the individual photoemission measurements, as in Fig. 2c. After compensation of momentum streaking, the photoemission intensity is integrated in two momentum regions with $k_y < 0$ and $k_y > 0$, respectively, and normalized by the total number of carriers in the two momentum regions at equilibrium. Subtracting the time-dependent number of counts with positive and negative $k_y$ components, we obtain a dimensionless value proportional to the microscopic current. To reveal the maximum amplitude of the current density, this procedure is performed for all carriers above the Dirac point ($E - E_D \geq 0$) within the momentum region $|k_x - K| < 0.15$ Å$^{-1}$. The result is then calibrated to physical values by multiplying with the carrier density in the conduction band of graphene with a Fermi level located 200 meV above the Dirac point, which can be computed from the density of states[27]. This yields a maximum current density of $j(t) = 38.2$ A/cm. This value is remarkably high, as it is even larger than in previous studies on topological insulators[8,19] where the scattering times were two orders of magnitude longer. Given the higher Fermi velocity and field amplitude, a higher current could be expected here. Yet, in a Drude-like regime this would not be sufficient to compensate for the much higher scattering rate. This confirms that in ballistic regimes the current does not follow the Drude model.

To extract the relative timing between the microscopic current in the Dirac cone, $j(t)$, and the electric field waveform, $E(t)$, we apply a phase delay analysis. We compute the product between the Fourier transform of the field waveform, $E(\omega) = F\{E(t)\}$, and the complex conjugate of the Fourier transform of the current, $j(\omega) = F\{j(t)\}$:

$$P(\omega) = E(\omega) \cdot J^\dagger(\omega).$$

The squared modulus of $P(\omega)$ peaks at the frequency of the pump field, while its phase corresponds to the timing difference (or phase delay) between the two oscillating quantities. We thus define the average phase difference as:

$$\phi = \frac{\sum_\omega |P(\omega)| \cdot \angle P(\omega)}{\sum_\omega |P(\omega)|},$$

where $\angle P(\omega)$ is the complex phase of $P(\omega)$ and summations only consider frequencies where $|P(\omega)|$ is larger than 50% of its peak value. To compute the phase delay, $\tau_i$, we need to divide this quantity by the central frequency of the pump field. We then obtain the mean phase delay, $\tau_\phi$, and the associated standard deviation of the mean over repeated measurements as:

$$\tau_\phi = \frac{\sum_i \tau_i}{N}, \quad \sigma_{\tau_\phi} = \frac{\sqrt{\sum_i [\tau_i - \tau_\phi]^2}}{N},$$

where $i = 1, \ldots, N$ runs over the different measurements.

**Semiconductor Bloch equations**

To simulate the time-dependent MIR-driven electron dynamics in graphene, we numerically solve the semiconductor Bloch equations[6,54,55] for a two-band system:

$$\hbar \frac{\partial f_e}{\partial t}(\mathbf{k}, t) = 2\hbar \mathbf{E}(t) \cdot \Im\{\mathbf{d}_{he}(\mathbf{k}) \, p_{eh}(\mathbf{k}, t)\} - e\mathbf{E}(t) \cdot \nabla_k f_e(\mathbf{k}, t) + \Gamma_k^e$$

$$\hbar \frac{\partial f_h}{\partial t}(\mathbf{k}, t) = -2\hbar \mathbf{E}(t) \cdot \Im\{\mathbf{d}_{he}(\mathbf{k}) \, p_{eh}(\mathbf{k}, t)\} - e\mathbf{E}(t) \cdot \nabla_k f_h(\mathbf{k}, t) + \Gamma_k^h$$

$$i\hbar \frac{\partial p_{eh}}{\partial t}(\mathbf{k}, t) = \left[\varepsilon_e(\mathbf{k}) - \varepsilon_h(\mathbf{k}) - \frac{i\hbar}{T_2}\right] p_{eh}(\mathbf{k}, t) + \mathbf{E}(t) \cdot \mathbf{d}_{eh}(\mathbf{k})[f_e(\mathbf{k}, t) - f_h(\mathbf{k}, t)]$$
$$+ \mathbf{E}(t) \cdot [\mathbf{d}_{hh}(\mathbf{k}) - \mathbf{d}_{ee}(\mathbf{k})] p_{eh}(\mathbf{k}, t) - ie\mathbf{E}(t) \cdot \nabla_k p_{eh}(\mathbf{k}, t)$$

where $\mathbf{k}$ is the crystal momentum, $t$ is the time, $\hbar$ the reduced Planck constant, and $e$ the elementary charge. $f_e(\mathbf{k},t)$ and $f_h(\mathbf{k},t)$ are the electron distribution in the conduction and valence bands, respectively, and $p_{eh}(\mathbf{k},t)$ is the microscopic interband polarization. In these equations, $\mathbf{E}(t)$ is the MIR electric field, $T_2 = 5$ fs is the phenomenological dephasing time of the microscopic polarization, $\varepsilon_e(\mathbf{k})$ and $\varepsilon_h(\mathbf{k})$ are the energy values of the conduction and valence bands, and

$$\mathbf{d}(\mathbf{k}) = \begin{bmatrix} \mathbf{d}_{ee}(\mathbf{k}) & \mathbf{d}_{eh}(\mathbf{k}) \\ \mathbf{d}_{he}(\mathbf{k}) & \mathbf{d}_{hh}(\mathbf{k}) \end{bmatrix} = e \begin{bmatrix} \langle \Psi_e | i\nabla | \Psi_e \rangle & \langle \Psi_h | i\nabla | \Psi_e \rangle \\ \langle \Psi_e | i\nabla | \Psi_h \rangle & \langle \Psi_h | i\nabla | \Psi_h \rangle \end{bmatrix}$$

are the dipole moments. $\Gamma_k^e$ and $\Gamma_k^h$ are the time- and momentum-dependent scattering rates for electrons and holes, respectively, and will be discussed in the next section. In simulations without interband coupling (Fig. 4 and Extended Data Fig. 4), we set $\mathbf{d}_{eh}(\mathbf{k}) = \mathbf{d}_{he}^\dagger(\mathbf{k}) = 0$.

In the specific case of graphene, the Hamiltonian of a two-level system can be written as[27]:

$$H(\mathbf{k}) = \begin{bmatrix} 0 & \gamma h(\mathbf{k}) \\ \gamma h^*(\mathbf{k}) & 0 \end{bmatrix},$$

with

$$h(\mathbf{k}) = |h(\mathbf{k})|e^{i\varphi_k} = e^{i\frac{ak_x}{\sqrt{3}}} + 2e^{-i\frac{ak_x}{2\sqrt{3}}}\cos\left(\frac{ak_y}{2}\right)$$

and $a = -2.46$ Å is the nearest-neighbour lattice distance, $v_F = 1.072 \times 10^6$ m/s is the experimental value of the Fermi velocity and $\gamma = -2\hbar v_F/(\sqrt{3}a)$ is the hopping integral. From the Hamiltonian, it is then possible to compute the equilibrium eigenenergies and eigenfunctions:

$$\varepsilon_{e,h}(\mathbf{k}) = \pm\gamma|h(\mathbf{k})|, \qquad \Psi_{e,h}(\mathbf{k}) = \frac{1}{\sqrt{2}}\begin{bmatrix} e^{i\varphi_k} \\ \pm 1 \end{bmatrix}.$$

The out-of-equilibrium dynamics were simulated on a 250 × 250 2D square $(k_x, k_y)$ grid in momentum space. The grid is centered on one of the K points with coordinates $k_{x,K} = 1.703$ Å$^{-1}$ and $k_{y,K} = 0$, and it has a side length of 0.73 Å$^{-1}$. The simulated time axis ranges from $-100$ fs to 300 fs, with a step size of 2 fs. The electric field, $\mathbf{E}(t)$, is parallel to $k_y$ and its temporal profile is analytically fitted to the experimentally measured field. At the initial time instant ($t_0 = -100$ fs), we assume $p_{eh}(\mathbf{k},t_0) = 0$, while the electron distribution corresponds to a Fermi-Dirac distribution. The position of the Fermi level compared to the Dirac crossing is set to $\varepsilon_F = 0.2$ eV to match the experiment.

**Modelling of the scattering rates**

In simulations including scattering, the grid was adjusted to 50 × 50 while maintaining the size in momentum space, to save resources. To reproduce the time-dependent electron distribution observed in momentum space, it is imperative to realistically model the scattering rates:

$$\Gamma_k^e = \left.\frac{\partial f_e(\mathbf{k},t)}{\partial t}\right|_{coll}, \qquad \Gamma_k^h = \left.\frac{\partial f_h(\mathbf{k},t)}{\partial t}\right|_{coll}.$$

On these time scales, two physical phenomena mainly contribute to the scattering rates:

$$\Gamma_k^\alpha = \Gamma_{k,ep}^\alpha + \Gamma_{k,ee}^\alpha, \quad \alpha \in \{e,h\},$$

where $\Gamma_{k,ep}^\alpha$ and $\Gamma_{k,ee}^\alpha$ are, respectively, the electron-phonon and electron-electron scattering rates.

Starting from the electron-phonon interaction, the scattering rate can be written as[56]:

$$\Gamma_{k,ep}^\alpha = \sum_{v,\alpha',\mathbf{k}'} \{S_{\alpha'\alpha}^v(\mathbf{k}',\mathbf{k}) f_{\alpha'}(\mathbf{k}',t)[1-f_\alpha(\mathbf{k},t)] - S_{\alpha\alpha'}^v(\mathbf{k},\mathbf{k}') f_\alpha(\mathbf{k},t)[1-f_{\alpha'}(\mathbf{k}',t)]\},$$

where $v$ is the phonon mode considered and $\alpha, \alpha' \in \{e,h\}$ are the band indexes. The first term thus describes transitions from an occupied $(\mathbf{k}',\alpha')$ state to an empty $(\mathbf{k},\alpha)$ state due to the interaction with a $v$ phonon, while the second term describes the opposite process. While in the former the state $(\mathbf{k},\alpha)$ is empty and becomes occupied (thus, the rate is positive), in the latter the rate must be negative. The probability of these transitions is associated with the electron-phonon transition rate:

$$S^v_{\alpha'\alpha}(\boldsymbol{k}',\boldsymbol{k}) = \frac{2\pi}{\hbar}\left|G^v_{\alpha'\alpha}(\boldsymbol{k}',\boldsymbol{k})\right|^2$$
$$\cdot \left\{(n^v_q + 1)\,\delta[\varepsilon_\alpha(\boldsymbol{k}) - \varepsilon_{\alpha'}(\boldsymbol{k}') + \hbar\omega^v_q]\,\delta_{\boldsymbol{k}'=\boldsymbol{k}-\boldsymbol{q}} \right.$$
$$\left. + n^v_q\,\delta[\varepsilon_\alpha(\boldsymbol{k}) - \varepsilon_{\alpha'}(\boldsymbol{k}') - \hbar\omega^v_q]\,\delta_{\boldsymbol{k}'=\boldsymbol{k}+\boldsymbol{q}}\right\},$$

where $G^v_{\alpha'\alpha}(\boldsymbol{k}',\boldsymbol{k})$ is the electron-phonon coupling constant, $\delta$ is the Kronecker delta imposing energy and momentum conservation, $\boldsymbol{q}$ and $\omega^v_q$ are the momentum and angular frequency of a $v$ phonon, and $n^v_q$ is the number of $v$ phonons with a given frequency and momentum, which is described by the Bose-Einstein distribution.

To simplify our description, we assume that (i) Umklapp processes are negligible, and (ii) the number of phonons $n^v_q$ is constant and equal to its value at equilibrium (phonon bath). Moreover, previous works elucidated the dominating contribution of Γ longitudinal (LO) and transverse optical (TO) phonons[33,46]. We will thus limit our model to LO and TO phonons and, since they have similar phonon energies ($\omega^{LO}_q \simeq \omega^{TO}_q \simeq 164.6\,\hbar^{-1}$ meV) and matrix elements, but their angular dependencies compensate each other, we will treat them together with an effective electron-optical phonon coupling constant[56]:

$$\left|G^{LO}_{\alpha'\alpha}(\boldsymbol{k}',\boldsymbol{k}) + G^{TO}_{\alpha'\alpha}(\boldsymbol{k}',\boldsymbol{k})\right|^2 = \left|G^{OP}_{\alpha'\alpha}\right|^2 = 1.542\text{ eV}^2.$$

The effective matrix element is constant in momentum space, and its value is adjusted to reproduce the experimental data. Due to the isotropic angular dependence of this term, electron-optical phonon scattering will lead to a rapid spreading of the electron distribution around the Dirac cone.

In the same spirit, the scattering rate for electron-electron interaction can be written as[56]:

$$\Gamma^\alpha_{k,ee} = \sum_{\boldsymbol{k}_*,\boldsymbol{k}',\boldsymbol{k}'_*}\sum_{\alpha',\beta,\beta'}\left\{S_{ee}(\boldsymbol{k}',\boldsymbol{k}'_*,\boldsymbol{k},\boldsymbol{k}_*)\,f_{\alpha'}(\boldsymbol{k}')\,f_{\beta'}(\boldsymbol{k}'_*)\,[1 - f_\alpha(\boldsymbol{k})]\,[1 - f_\beta(\boldsymbol{k}_*)] \right.$$
$$\left. - S_{ee}(\boldsymbol{k},\boldsymbol{k}_*,\boldsymbol{k}',\boldsymbol{k}'_*)\,f_\alpha(\boldsymbol{k})\,f_\beta(\boldsymbol{k}_*)\,[1 - f_{\alpha'}(\boldsymbol{k}')]\,[1 - f_{\beta'}(\boldsymbol{k}'_*)]\right\}\delta_{\boldsymbol{k}+\boldsymbol{k}_*=\boldsymbol{k}'+\boldsymbol{k}'_*},$$

where $\alpha,\alpha',\beta,\beta' \in \{e, h\}$ are the band indices. The first term, therefore, describes the scattering between a pair of particles in the initial states $(\alpha',\boldsymbol{k}')$ and $(\beta',\boldsymbol{k}'_*)$ to the final states $(\alpha,\boldsymbol{k})$ and $(\beta,\boldsymbol{k}_*)$, and the second term refers to the opposite process. The associated transition rate can thus be written as[56]:

$$S_{ee}(\boldsymbol{k},\boldsymbol{k}_*,\boldsymbol{k}',\boldsymbol{k}'_*) = \left(\frac{2\pi}{\hbar}\right)^2 |M|^2\,\delta[\varepsilon_{\beta'}(\boldsymbol{k}'_*) + \varepsilon_{\alpha'}(\boldsymbol{k}') - \varepsilon_\beta(\boldsymbol{k}_*) - \varepsilon_\alpha(\boldsymbol{k})],$$

where the matrix element is:

$$|M|^2 = \frac{1}{2}[|V(q)|^2 + |V(q')|^2 - V(q)\,V(q')]$$

and $q = |\boldsymbol{k} - \boldsymbol{k}'|$ and $q' = |\boldsymbol{k} - \boldsymbol{k}'_*|$. The Coulomb potential between electron pairs is:

$$V(q) = \frac{2\pi e^2 V_{\text{eff}}}{\epsilon(q)\,qA}\frac{1 + \cos\phi_{\boldsymbol{k}\boldsymbol{k}'}}{2}\frac{1 + \cos\phi_{\boldsymbol{k}_*\boldsymbol{k}'_*}}{2}$$

$$V(q') = \frac{2\pi e^2 V_{\text{eff}}}{\epsilon(q')\, q'A} \frac{1 + \cos\phi_{\boldsymbol{k}\boldsymbol{k}'_*}}{2} \frac{1 + \cos\phi_{\boldsymbol{k}_*\boldsymbol{k}'}}{2}$$

and $A = 4\pi^2/\Delta k^2$. The screening of the Coulomb interaction is given by the dielectric function $\epsilon(q) = 1 + v_c(q)\, D(\varepsilon_F)\, \tilde{\Pi}(q)$[56,57], where:

$$v_c(q) = \frac{2\pi e^2}{\kappa q}, \quad \kappa = \frac{e^2}{r_S \hbar v_F}, \quad r_S = \frac{\hbar v_F}{E_F}, \quad D(\varepsilon) = \frac{|\varepsilon|}{2\pi \hbar^2 v_F^2}$$

$$\tilde{\Pi}(q) = \begin{cases} 1, & q < 2k_F \\ 1 + \dfrac{\pi q}{8 k_F} - \dfrac{\sqrt{q^2 - 4 k_F^2}}{2q} - \dfrac{q}{4 k_F}\arcsin\left(\dfrac{2 k_F}{q}\right), & \text{otherwise} \end{cases}$$

and $r_S$ is the Wigner-Seitz radius of graphene, $k_F = \varepsilon_F/(\hbar v_F)$ the Fermi wavevector, and $D(\varepsilon)$ the density of states close to the Dirac point. This approximation holds for $n \geq n_{th} = 10^{12}$ cm$^{-2}$, thus if $\varepsilon_F \geq$ 37.14 meV, as in our case.

In this form, the scattering rate for electron-electron interaction is computationally extremely expensive. Therefore, we assume that the electron distribution for the second particle can be approximated by the equilibrium Fermi-Dirac distribution of the carriers:

$$f_\beta(\boldsymbol{k}_*) \simeq f_\beta^{(0)}(\boldsymbol{k}_*), \quad f_{\beta'}(\boldsymbol{k}'_*) \simeq f_{\beta'}^{(0)}(\boldsymbol{k}'_*).$$

This allows us to write a simplified expression for the electron-electron scattering rate:

$$\Gamma_{k,ee}^\alpha \simeq \sum_{\boldsymbol{k}'} \sum_{\alpha',\beta,\beta'} \left\{ S_{ee,\beta\beta'}^{\text{eff},1}(\boldsymbol{k},\boldsymbol{k}')\, f_{\alpha'}(\boldsymbol{k}')[1 - f_\alpha(\boldsymbol{k})] - S_{ee,\beta\beta'}^{\text{eff},2}(\boldsymbol{k},\boldsymbol{k}')\, f_\alpha(\boldsymbol{k})[1 - f_{\alpha'}(\boldsymbol{k}')] \right\},$$

where we use that $S_{ee}(\boldsymbol{k},\boldsymbol{k}_*,\boldsymbol{k}',\boldsymbol{k}'_*) = S_{ee}(\boldsymbol{k}',\boldsymbol{k}'_*,\boldsymbol{k},\boldsymbol{k}_*)$ and define:

$$S_{ee,\beta\beta'}^{\text{eff},1}(\boldsymbol{k},\boldsymbol{k}') = \sum_{\boldsymbol{k}_*,\boldsymbol{k}'_*} S_{ee}(\boldsymbol{k},\boldsymbol{k}_*,\boldsymbol{k}',\boldsymbol{k}'_*)\, f_{\beta'}^{(0)}(\boldsymbol{k}'_*)\left[1 - f_\beta^{(0)}(\boldsymbol{k}_*)\right] \delta_{\boldsymbol{k}+\boldsymbol{k}_* = \boldsymbol{k}'+\boldsymbol{k}'_*}$$

$$S_{ee,\beta\beta'}^{\text{eff},2}(\boldsymbol{k},\boldsymbol{k}') = \sum_{\boldsymbol{k}_*,\boldsymbol{k}'_*} S_{ee}(\boldsymbol{k},\boldsymbol{k}_*,\boldsymbol{k}',\boldsymbol{k}'_*)\, f_\beta^{(0)}(\boldsymbol{k}_*)\left[1 - f_{\beta'}^{(0)}(\boldsymbol{k}'_*)\right] \delta_{\boldsymbol{k}+\boldsymbol{k}_* = \boldsymbol{k}'+\boldsymbol{k}'_*}.$$

In our simulations, we only focus on a small momentum space region around one of the six equivalent K points. In doing so, we neglect intervalley electron-phonon and electron-electron scattering processes, which are not expected to contribute significantly in these conditions[33]. However, for electron-electron scattering, it is possible that an electron scatters from one state to another within the same valley by interacting with a charge carrier in a different one. For this reason, we effectively account for these processes by multiplying $S_{ee,\beta\beta'}^{\text{eff},1}$ and $S_{ee,\beta\beta'}^{\text{eff},2}$ by the number of equivalent K valleys, i.e. six. To reproduce the experimental data, we assumed $V_{\text{eff}} = 4.386$. It is important to notice that the dependence on the relative angle between $\boldsymbol{k}$, $\boldsymbol{k}'$, $\boldsymbol{k}_*$, and $\boldsymbol{k}'_*$ determines the characteristic angular dependence of electron-electron scattering in graphene[46], also termed collinear scattering[40].

**Relaxation time approximation**

To reduce the computational costs, a customary approach when solving the semiconductor Bloch equations is to invoke the relaxation time approximation[6,19,54,55]. In this context, the scattering rate $\Gamma_k^\alpha$ is assumed to be inversely proportional to a time constant, $T_1$, associated with the mean free time between scattering events. Defining $f_\alpha^{(0)}(\mathbf{k})$ the equilibrium electron distribution in the $\alpha$ band, we obtain:

$$\Gamma_k^\alpha = \left.\frac{\partial f_\alpha(\mathbf{k},t)}{\partial t}\right|_{\text{coll}} = -\frac{f_\alpha(\mathbf{k},t) - f_\alpha^{(0)}(\mathbf{k})}{T_1}.$$

It is important to notice that, even though this approach allows to extract an effective time constant, it is not enough to reproduce the experimental data. In fact, in this approximation the scattering rate at a given point $\mathbf{k}$ in momentum space will only depend on the instantaneous electron distribution at that point, and not on any other point in momentum space. This makes it impossible to reproduce any carrier redistribution dynamics in momentum space due to, e.g., the interaction with optical phonons.

**Injection time analysis**

For an empty upper Dirac cone, one expects the carrier injection rate due to LZM tunnelling to closely follow the magnitude of the electric field and the electron population to increase stepwise, with the highest injection rates occurring at the largest absolute field values. In an *n*-doped system and for high injection rates, as in the present experiment, Pauli blocking modifies this simple picture (Extended Data Fig. 5a). Electrons can only be injected, when there is sufficient empty space in the CB near the K point. Consequently, intraband transport plays a decisive role in determining the injection timing. Carrier injection happens most efficiently during the intervals between the extrema of the electric field and the carrier displacement. Our simulations clearly show that this results in a phase delay between the population jumps and the field, which grows as more population has been transferred. Beyond the effects of Pauli blocking and scattering, the injection dynamics can strongly be influenced by quantum inferences[58]. Here we find that they can cause a deviation of a step-like increase in the CB population (or electron distribution width) by transforming the plateau regions, following a strong injection phase, into local minima through destructive interference of electron wave packets (Extended Data Fig. 7). While, the signal-to-noise ratio of the measured CB population, does not allow us to resolve the jumps in the population during carrier injection experimentally, the dynamics of the rise of the carrier density can be well reproduced with the theory.

The simulations also show that the elongation of the CB electron distribution in momentum space is closely tied to the carrier population (Extended Data Fig. 5b). Interestingly, the weak oscillations in the width of the electron distribution, which are superimposed on the stepwise increase, become more pronounced with scattering (Fig. 4c). This effect is probably caused by a shortening of the effective injection time window. In the absence of scattering, the electron displacement follows the vector potential and is therefore phase-shifted by $\pi/2$ relative to the electric field. With scattering, the phase shift between the electric field and the displacement of the electrons shrinks, such that periods of rapid population increase occur only a few femtoseconds before an extremum in the displacement (Extended Data Fig. 5). Experimentally, the oscillation of the width of the electron distribution at twice the frequency of the driving field is clearly visible during the first three cycles of the electric field, i.e., at times when there is still substantial carrier injection (Fig. 3c). The observed relative phase with respect to the electric field transient and the COM displacement agrees well with the simulations that include scattering. The data thus support the picture of a well that is periodically driven by the electric field and opened and closed by the oscillating carriers in the CB.

**Data Availability**

The data supporting the findings of this study are available from the corresponding authors upon request.

**Acknowledgements** We thank Prof. Dr. Jaroslav Fabian for helpful discussions. This work is supported by the Deutsche Forschungsgemeinschaft (DFG, German Research Foundation) through SFB 1083, project-ID 223848855, SFB 1277, project-ID 314695032, research grant HO 2295/9, research grant HU1598/8, GRK 2905, project-ID 502572516, DFG major instrumentation INST 89/520, project ID 445487514, and by the European Research Council (ERC) Synergy Grant, project-ID 101071259. G.I. and C.B. acknowledge financial support from the Alexander von Humboldt Foundation.

**Author Contributions** R.H., U.H., and F.S.T conceived the study. V.E., G.I., M.M., L. Münster, J.H., L. Machtl, and C.B. carried out the experiment. M.M., V.E., L. Münster, J.H., R.W., S.Z., S.I., L. Machtl, and J.G. realized the experimental setup. V.E. analysed the experimental data. H.Y., C.K., F.C.B., and F.S.T. prepared the samples. G.I. carried out the SBE calculations with support from V.E. All authors analysed and discussed the results. V.E., G.I., R.H., and U.H. wrote the paper with contributions from all authors.

**Author Information** Reprints and permissions information is available at www.nature.com/reprints. The authors declare no competing financial interests. Correspondence and requests for materials should be addressed to R. Huber (rupert.huber@ur.de) or U. Höfer (hoefer@physik.uni-marburg.de).

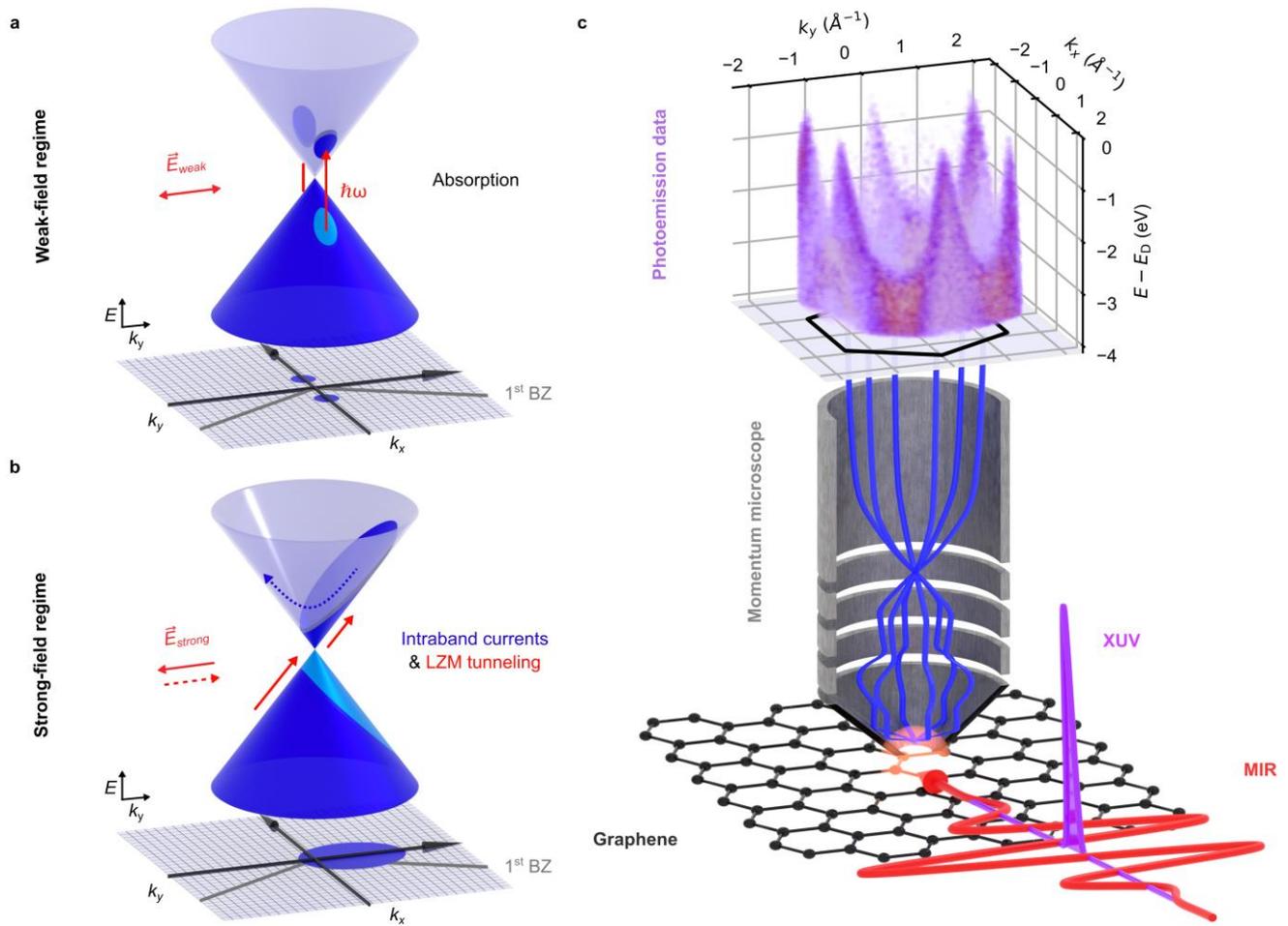

**Figure 1 | Landau-Zener-Majorana transitions in 2D momentum space. a,** Illustration of the fundamental electronic processes governing light-matter interaction in undoped graphene at the K point in the weak-field regime. Here, electrons can be promoted from the valence band (VB) to the conduction band (CB) by absorption. In graphene, the optical selection rules favour transitions in the momentum direction perpendicular to the polarization of the electric field. **b,** In the strong field regime, electrons can undergo ballistic acceleration by an electric field (dashed dark blue arrow). Simultaneously, the field can drive interband LZM transitions on subcycle timescales, allowing electrons to tunnel from the VB to the CB (solid red arrow). This results in an elongated electron distribution in the previously unoccupied CB, with the weight of the electron population in the direction parallel to the field. **c,** Experimental scheme (bottom half): an *s*-polarized, phase-stable MIR pulse (red) drives a monolayer of graphene, a *p*-polarized XUV probe pulse (purple) at variable time delay, *t*, provides snapshots of the induced dynamics with subcycle temporal resolution. The photoemitted electrons (blue) enter a photoemission momentum microscope. This enables the simultaneous detection of both parallel

momenta, $k_x$ and $k_y$, and kinetic energy for every individual electron. The upper half shows a femtosecond snapshot of the 2D band structure of a monolayer of graphene, recorded with the momentum microscope. The field of view extends over the entire first BZ, allowing to track the electronic occupation slightly beyond the six K points.

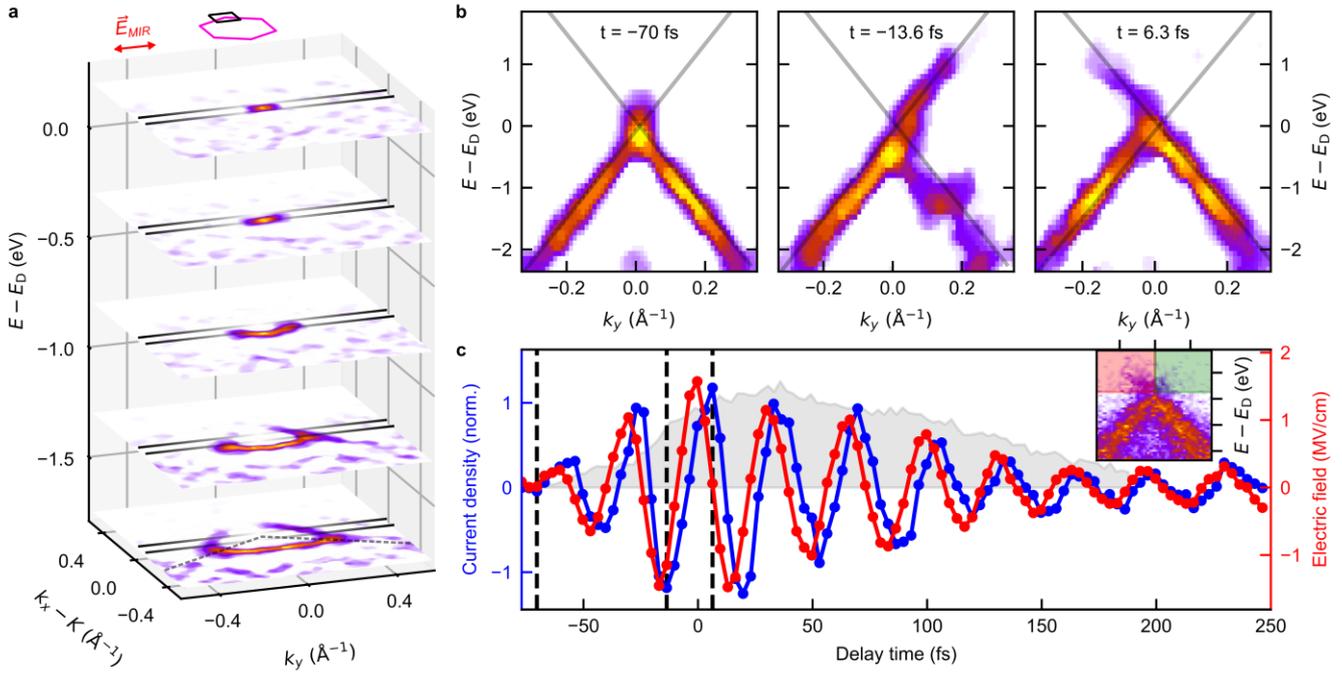

**Figure 2 | Lightwave-driven currents in graphene. a,** Zoomed-in curvature-filtered cuts through the band structure at a selected K point. The red arrow (top) indicates the MIR polarization. The dashed black line (bottom) marks the BZ border. The solid black lines represent the integration range in momentum space used in the three right panels. **b,** Streaking-compensated and curvature-filtered view into the band structure of graphene at the Dirac point before the arrival of the MIR pulse ($t = -70$ fs) and for selected delay times during the interaction with the MIR pulse, $t = -13.6$ fs and $t = 6.3$ fs (raw data, Extended Data Fig. 9). The latter two snapshots show how the electron distribution is accelerated along the Dirac cone away from the equilibrium Fermi level. **c,** Current extracted from occupation imbalance in the photoemission maps (blue points) and MIR waveform reconstructed from the momentum streaking (red points). Black dashed lines mark the temporal positions shown in **b**. The lightwave-driven current is extracted by tracking the imbalance in the occupation for positive and negative momenta (red and green shades areas) in the raw photoemission spectra (inset) and normalized on the total amount of carriers in both regions before the pulse arrives. The grey shaded area indicates the normalized population increase in the conduction band extracted from a similar measurement.

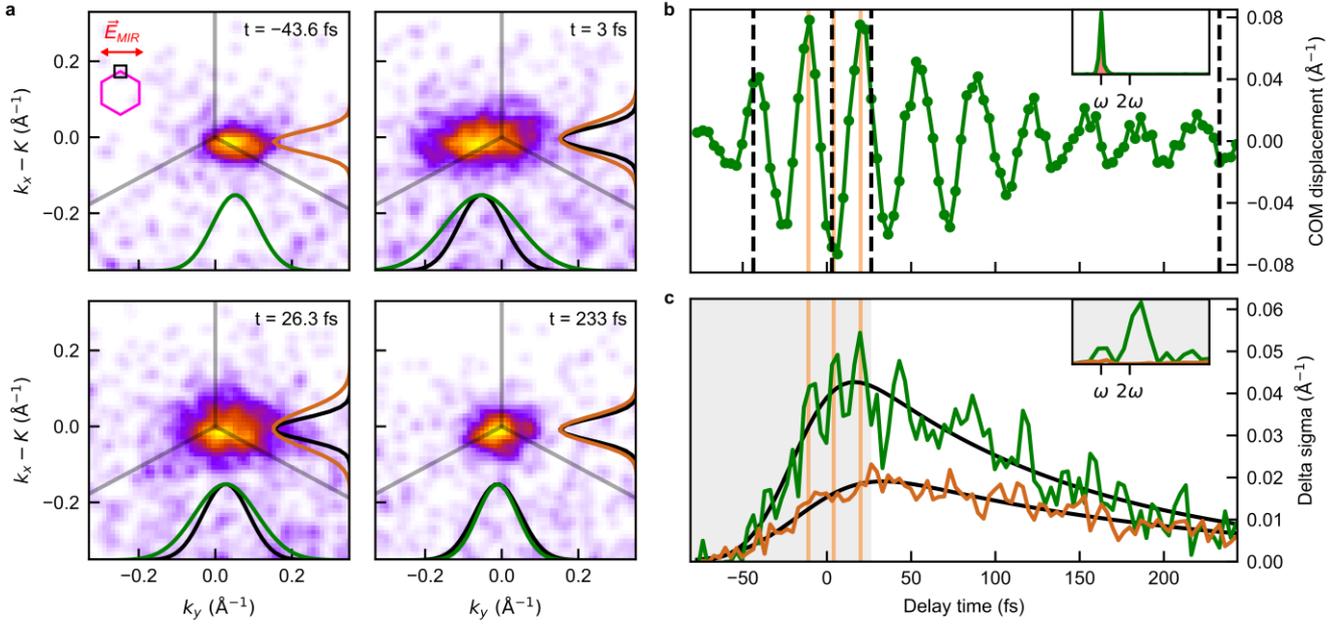

**Figure 3 | Subcycle electron dynamics in 2D momentum space. a,** 2D carrier distribution in the conduction band for selected delay times. The inset indicates the position of the evaluated window in momentum space, and the arrow shows the MIR polarization. The grey lines mark the BZ borders. The Gaussian fits illustrate the width of the projection of the carrier distribution on the $k_x$ axis (brown) and the $k_y$ axis (green). The widths of the first panel are shown in black in the later snapshots as a reference. Under the influence of the field, the carrier distribution is displaced along $k_y$ ($t = -43.6$ fs). Owing to LZM transitions, the carrier distribution is elongated along $k_y$ forming an ellipse ($t = 3$ fs). For later delays ($t = 26.3$ fs), scattering leads to a broadening also along $k_x$. For vanishing fields, the electron distribution relaxes back to its original shape ($t = 233$ fs). **b,** Displacement of the centre of mass of the electron distribution. The inset shows the Fourier transform of the oscillating electron distribution (green line) and the driving field (shaded red area). Dashed black lines mark the temporal positions of the momentum maps in **a**. **c,** Evolution of the change in the size of the electron distribution, characterised by the width of the Gaussians used to fit the carrier distribution along the $k_x$ axis (brown) and the $k_y$ axis (green). The Fourier transform in the inset reveals oscillations in the width of the distribution along the $k_y$ axis with pronounced components at the second harmonic of the driving frequency during the initial half-cycles of the driving field (shaded grey area). The yellow lines in **b** and **c** are guides to the eye to relate the oscillation in the width of the electron distribution to the motion of its centre of mass.

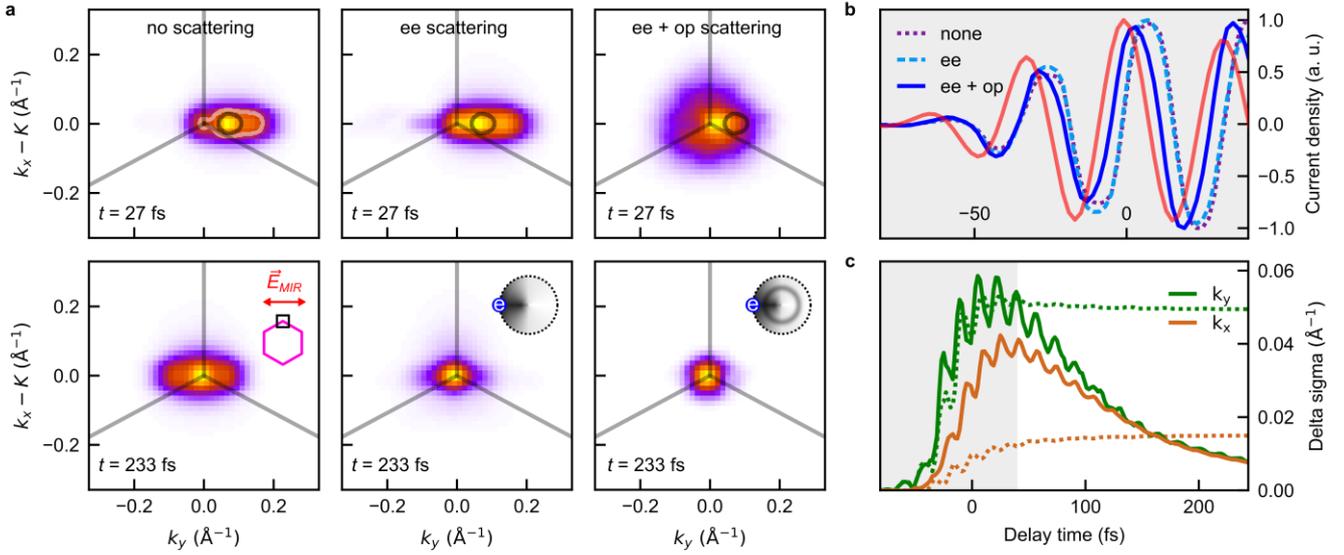

**Figure 4 | Ultrafast scattering channels in momentum space. a,** Simulated 2D carrier distribution in the conduction band for selected delay times with different scattering contributions. The leftmost panels show how the carrier distribution behaves for intermediate and late delay times $t$, based on the SBEs without scattering. The light grey line marks the half-maximum contour, highlighting the shape of the carrier distribution. For comparison, the dark grey line shows the corresponding contour when additionally excluding interband coupling, emphasizing its role in reproducing the experimental observations. It is included in the other panels for reference. The middle and right panels show the carrier distribution including different scattering channels: electron–electron (ee) scattering, and electron–electron combined with electron–optical phonon (ee + op) scattering, respectively. The cartoons in the insets illustrate the regions in momentum space accessible for an electron (small, blue sphere) via scattering (dark regions: accessible, bright regions: inaccessible). **b,** Time-dependent evolution of the extracted current for different scattering mechanisms (dotted dark blue line: no scattering; dotted light blue line: ee scattering; solid blue line: ee + op scattering), the driving field is shown as a red line. Electron-electron scattering affects the phase-delay only slightly, while including op scattering impacts it significantly. **c,** Temporal evolution of the change in the width of the electron distribution along $k_x$ (brown) and $k_y$ (green) for different levels of theory (dotted lines: no scattering; solid lines: ee + op scattering). Without scattering (dotted curve), the distribution is broadened strongly only in the direction of the field. By including scattering with optical phonons (solid curve), it is possible to reach a qualitative agreement with the experiment.

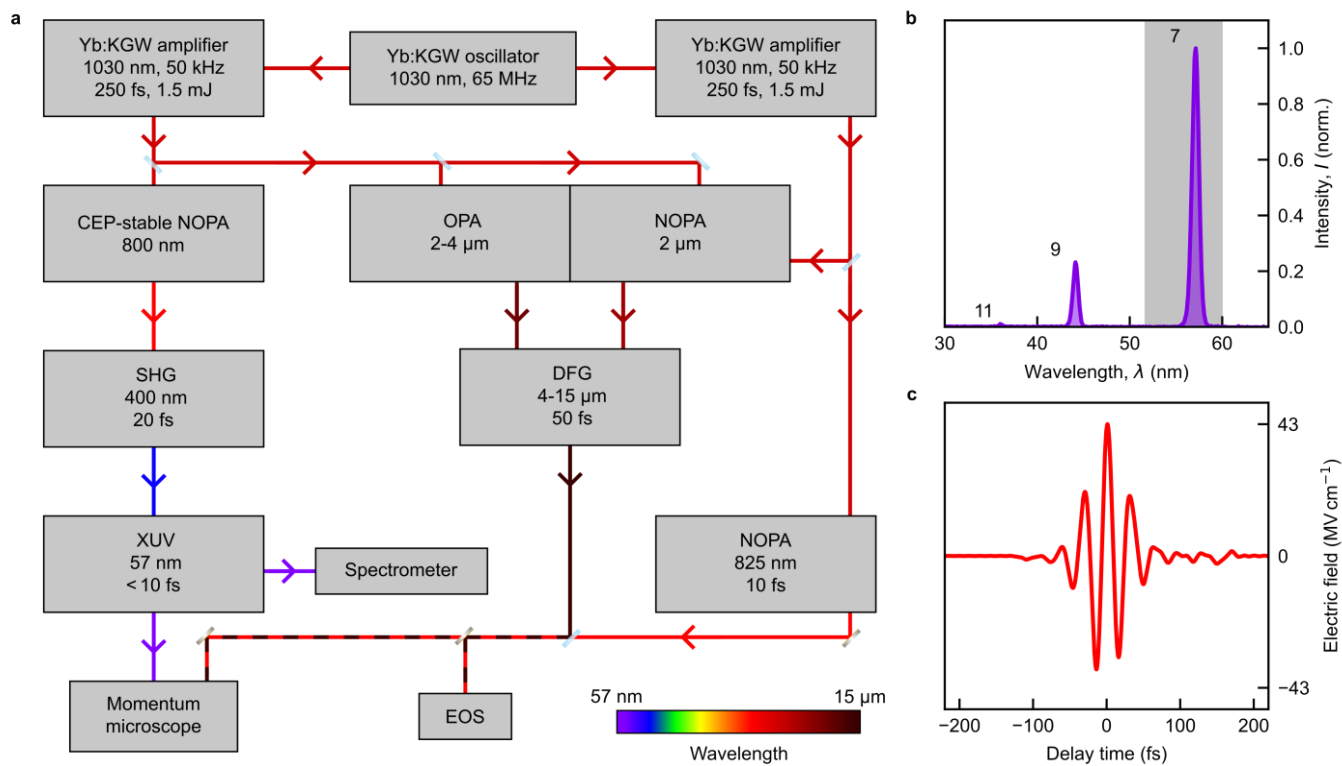

**Extended Data Figure 1 | Experimental setup. a,** A Yb:KGW oscillator seeds two Yb:KGW amplifiers, each providing a pulse energy of 1.5 mJ at a repetition rate of 50 kHz. Half of the output of one amplifier is upconverted to 400-nm pulses with a pulse duration of 20 fs, which are then used to generate sub-10-fs XUV pulses by HHG in argon (left side). In parallel, widely tuneable, intense, few-cycle and CEP-stable MIR pulses are generated by DFG of the outputs of a dual-branch OPA, which is supplied with the remaining half of the output of the first amplifiers (middle). The output of the second amplifier is used to almost exclusively pump the last amplification stage in the dual-branch OPA. Additionally, a NOPA generates 10-fs-long NIR pulses (right side), which can be used to detect the MIR waveforms (EOS). Finally, all beams are guided into the UHV chamber of the momentum microscope where they are spatially overlapped. **b,** Generated XUV spectrum. The intense 7th harmonic order centred around 57 nm is selected by multilayer mirrors (grey shaded area) and is used as a probe pulse in the subcycle momentum microscopy experiments. **c,** Typical MIR waveform measured by electro-optic sampling (EOS).

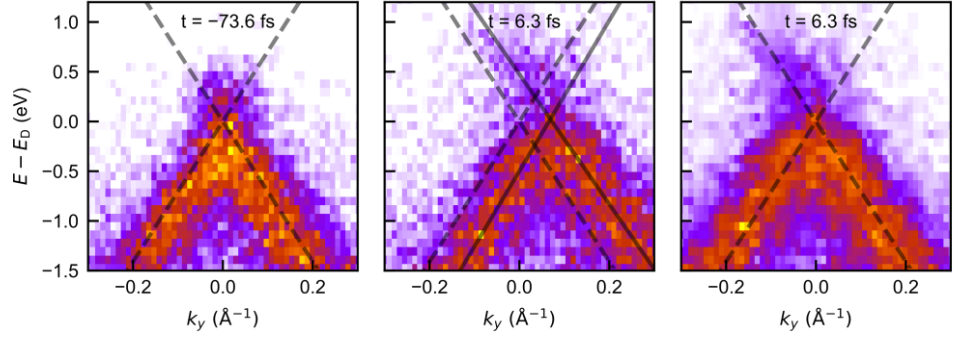

**Extended Data Figure 2 | Electric field extraction.** Photoemission spectra in different streaking conditions. The first two panels show a band structure snapshot taken in equilibrium (left panel) and a snapshot with strong streaking along $k_y$ by the *s*-polarized MIR field (middle panel). The electric field is reconstructed by matching the streaked band structures to the equilibrium case. Once the electric field is known, the streaking effects can be compensated (right panel). The black dashed lines are a guide to the eye representing the equilibrium band structure. Applying the calculated shifts along energy and momentum, according to the extracted field from the middle panel, to the black dashed line yields the solid black line.

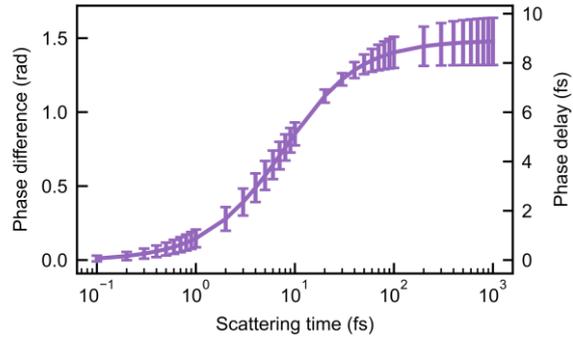

**Extended Data Figure 3 | Phase delay in the relaxation time approximation.** Simulations based on the SBEs in relaxation time approximation reveal how the phase delay between an applied waveform (central frequency, 30 THz) and the lightwave-driven current scales with the scattering time. For very small scattering times, the phase difference converges to zero, while for longer scattering times it converges to π/2. The strongest change in the phase delay can be observed for scattering times on the order of a half cycle of the driving waveform.

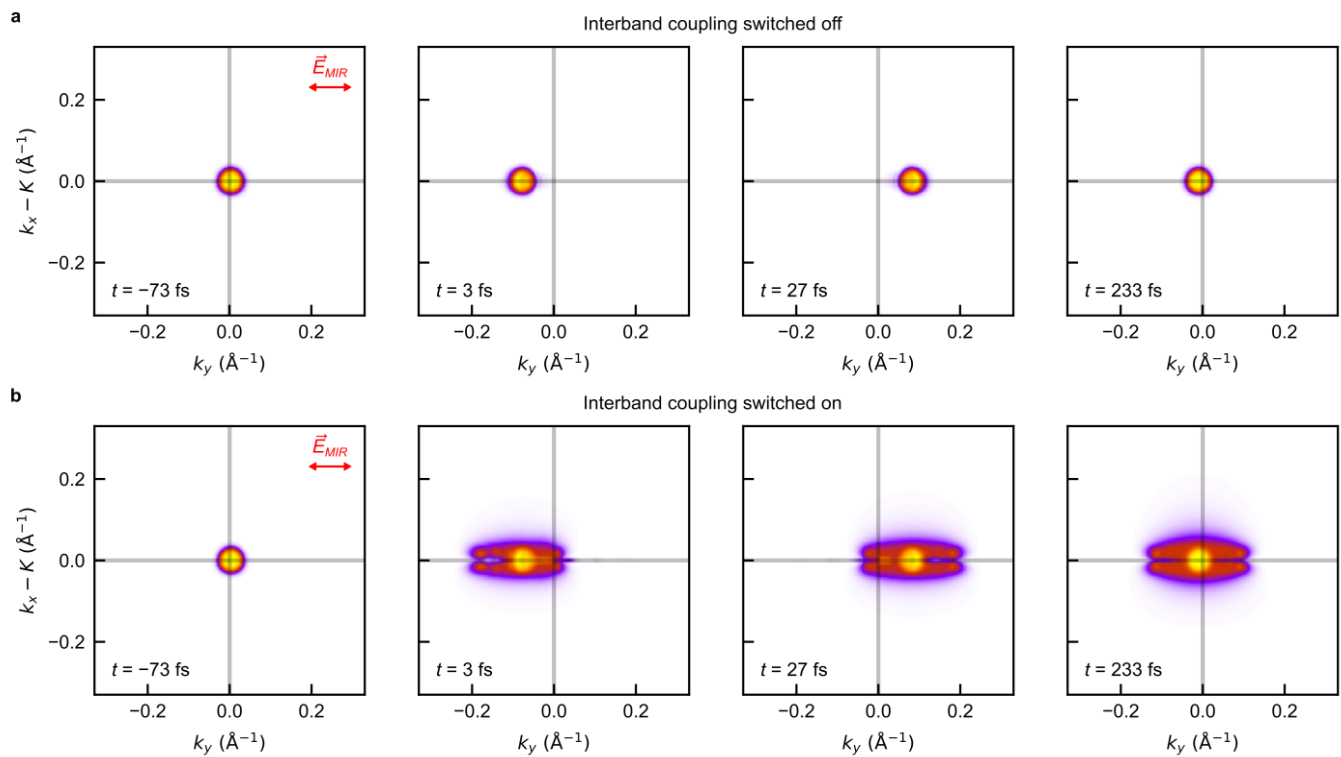

**Extended Data Figure 4 | Interband coupling in momentum space.** Simulated 2D carrier distribution in the conduction band for selected delay times without scattering and with interband coupling switched off, **a**, and switched on, **b**. The interband coupling causes a large elongation of the carrier distribution along the polarization of the driving light field, filling the gap between the displaced electrons distribution and the K point.

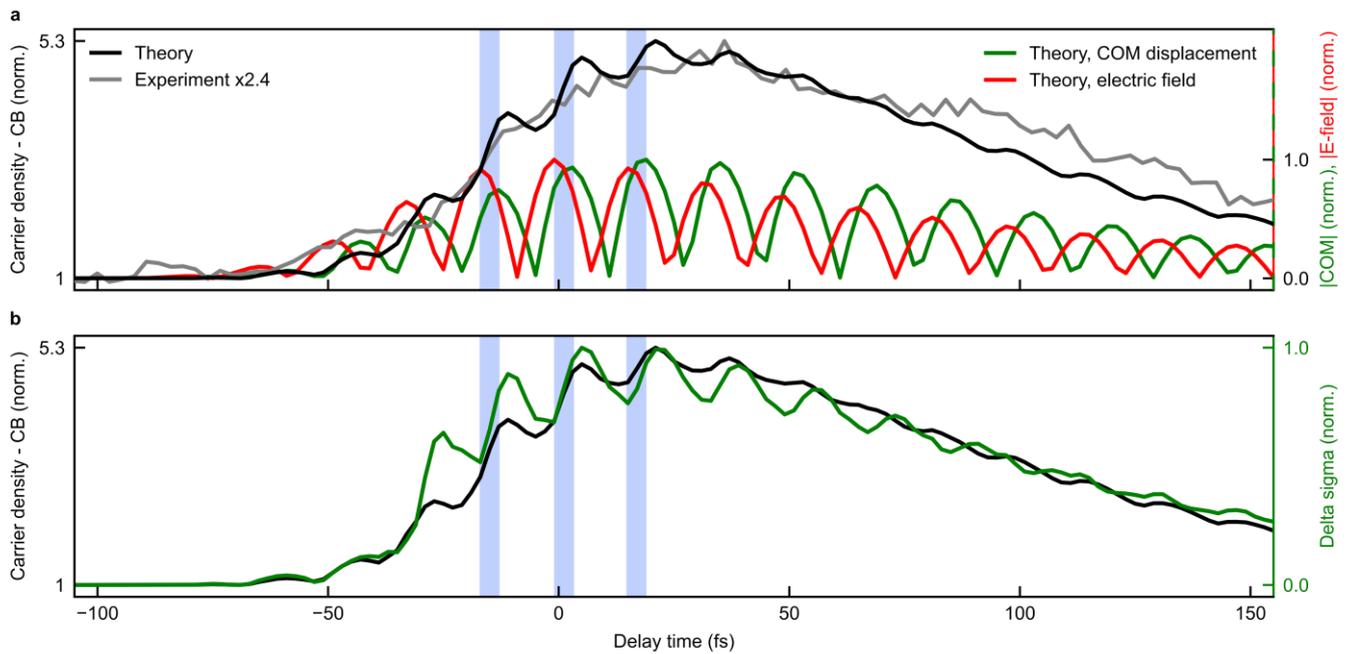

**Extended Data Figure 5 | Population dynamics in the conduction band. a,** Normalized carrier density in the conduction band extracted from our theory model with electron-phonon and electron-electron scattering (black curve) and from experimental data (grey curve). The observed difference in the number of additional carriers between theory and experiment is to be expected, as the theory does not account for inherent constraints of the measurement itself, such as matrix element effects. The population rises fastest during the time window between a maximum of the field (red curve) and the the centre-of-mass (COM) displacement (green curve). **b,** During the first few cycles, the evolution of the width of the electron distribution (green curve) closely follows the population dynamics (black curve).

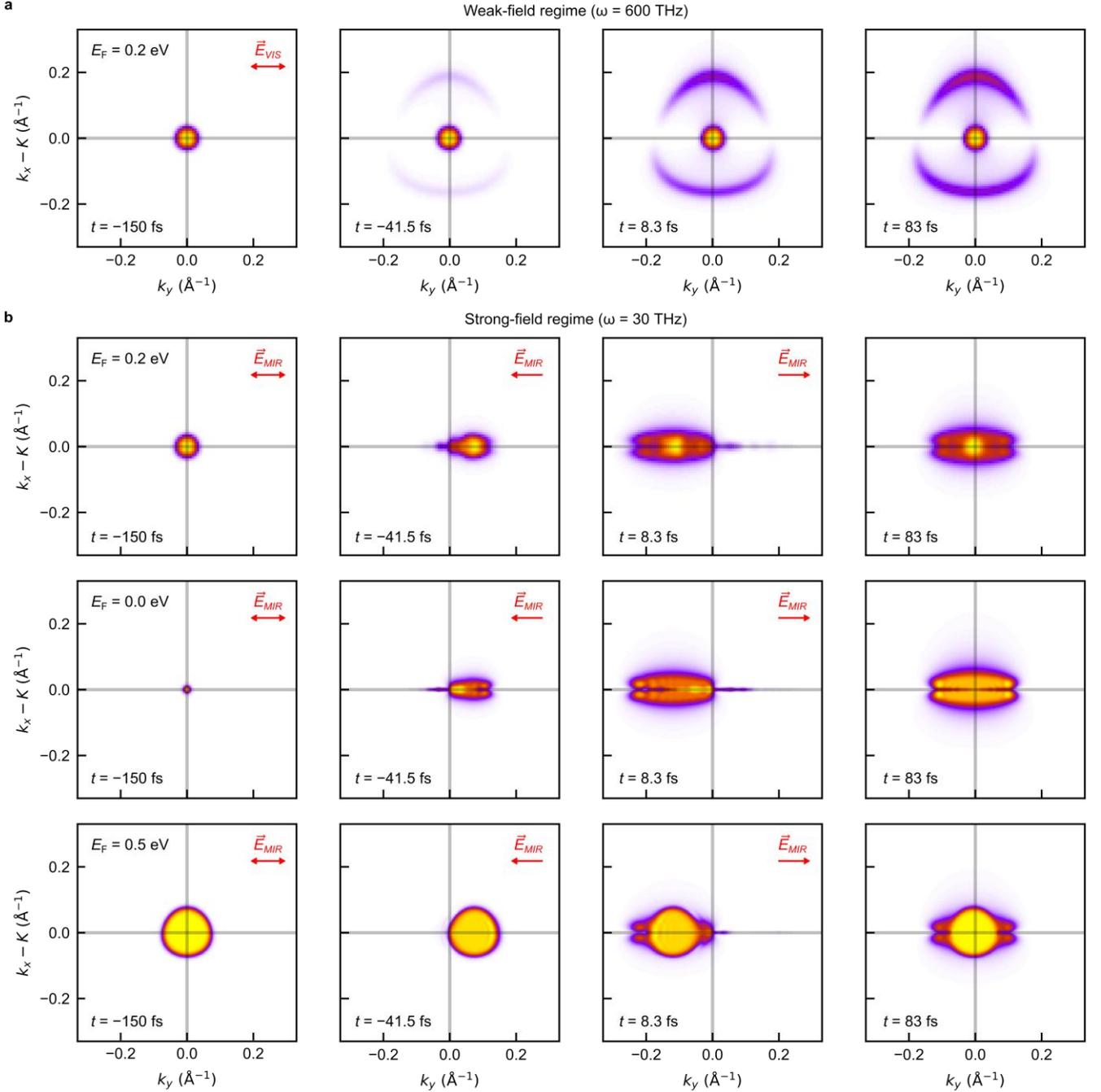

**Extended Data Figure 6 | Weak- and strong-field regime in momentum space.** Simulated 2D carrier distribution in the conduction band for selected delay times without scattering in the weak-field, **a**, and strong-field regime, **b**. The interaction in the weak field regime is dominated by absorption, which is strongly affected by the selection rules due to the pseudospin texture of graphene leading to a population increase in regions perpendicular to the polarization of the light pulse. Contrary to this, in the strong-field regime, the interaction with light is characterized by a unidirectional elongation of the electron distribution parallel to the electric field in addition to a displacement of the Fermi surface. Varying the doping highlights the difference between the intraband and interband signatures.

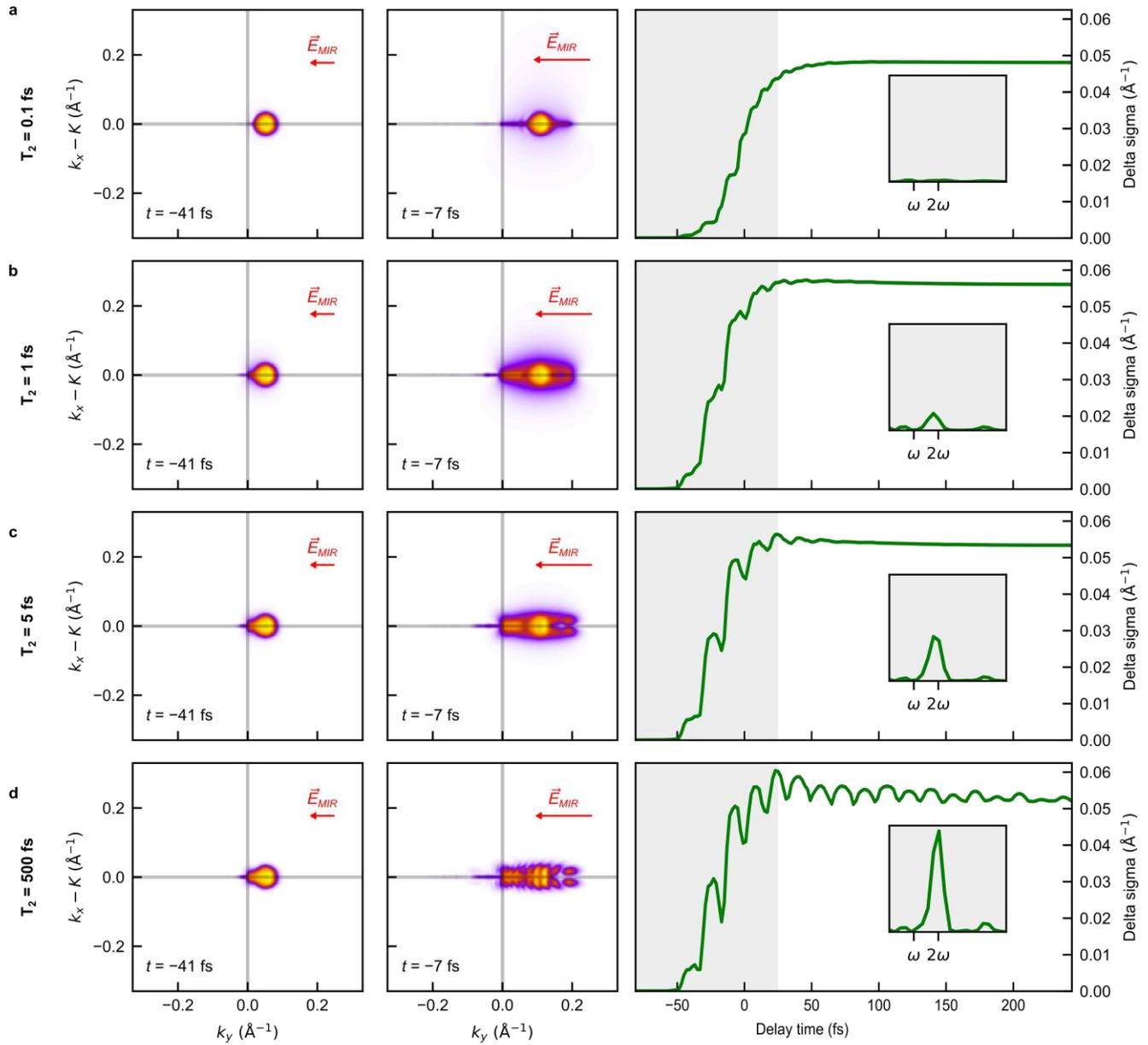

**Extended Data Figure 7 | Effect of dephasing in momentum space.** Simulated 2D carrier distribution in the conduction band for different dephasing times: $T_2 = 0.1$ fs, **a**, $T_2 = 1$ fs, **b**, $T_2 = 5$ fs, **c**, $T_2 = 500$ fs, **d**. The 2D carrier distribution is shown after the first (left panel) and second cycle (middle panel). For very fast dephasing, the efficiency for LZM transitions decreases drastically, manifesting in a lack of additional carriers in the second panel of **a**. For very slow dephasing, signatures of LZSM interference emerge after the second cycle (fishbone pattern in the second panel of **d**). They manifest in a modulation at the second harmonic of the driving field in the elongation of the electron distribution along the direction parallel to the electric field (right panel), as highlighted in the inset by its Fourier transform during the first half-cycles of the driving field (grey area). The scale of the vertical axis is fixed across the instets.

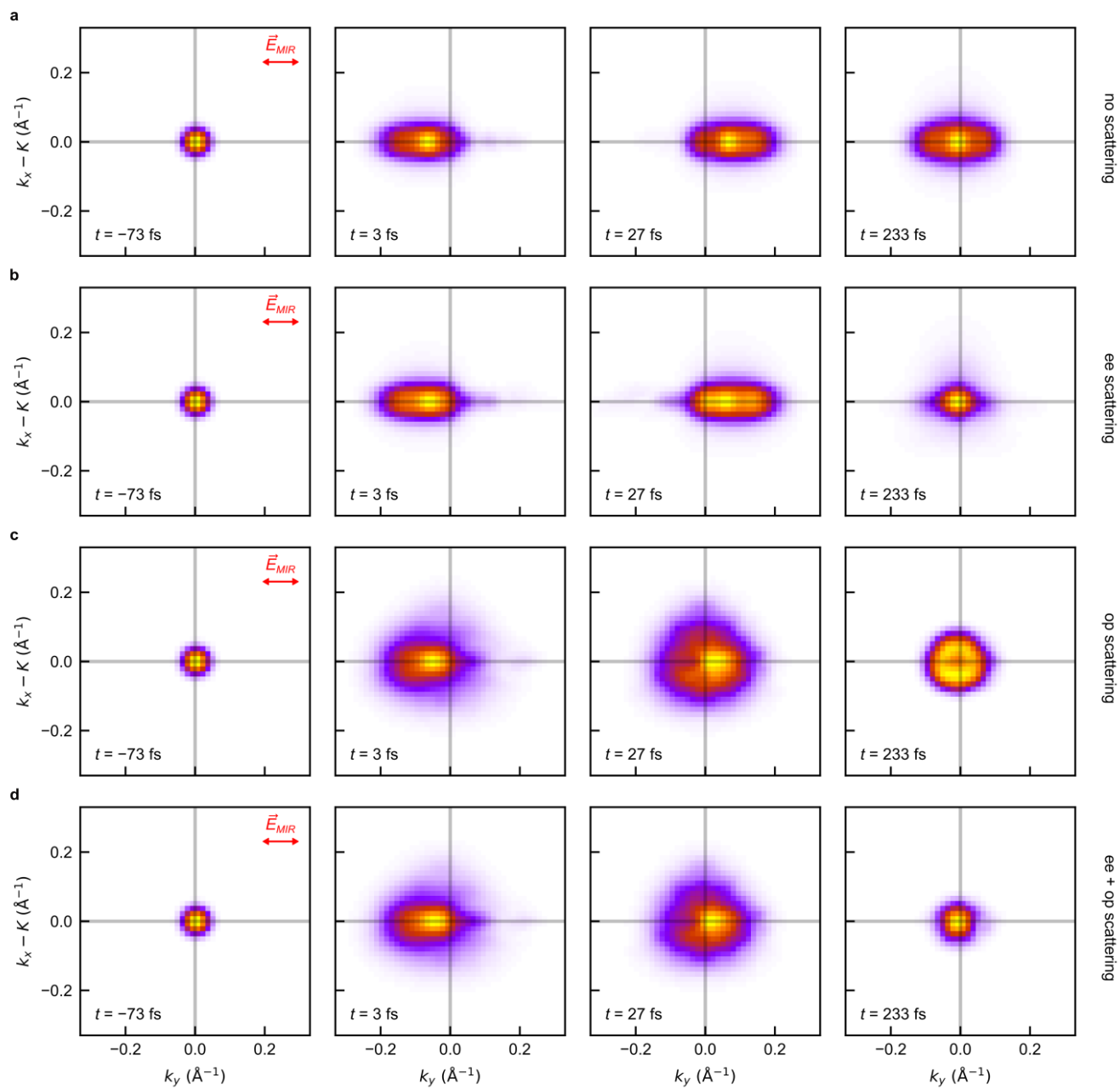

**Extended Data Figure 8 | Scattering channels in momentum space.** Simulated 2D carrier distribution in the conduction band for selected delay times with different scattering contributions: no scattering, **a**, only electron-electron scattering, **b**, only scattering with optical phonons, **c**, a combination of electron-electron scattering and scattering with optical phonons, **d**.

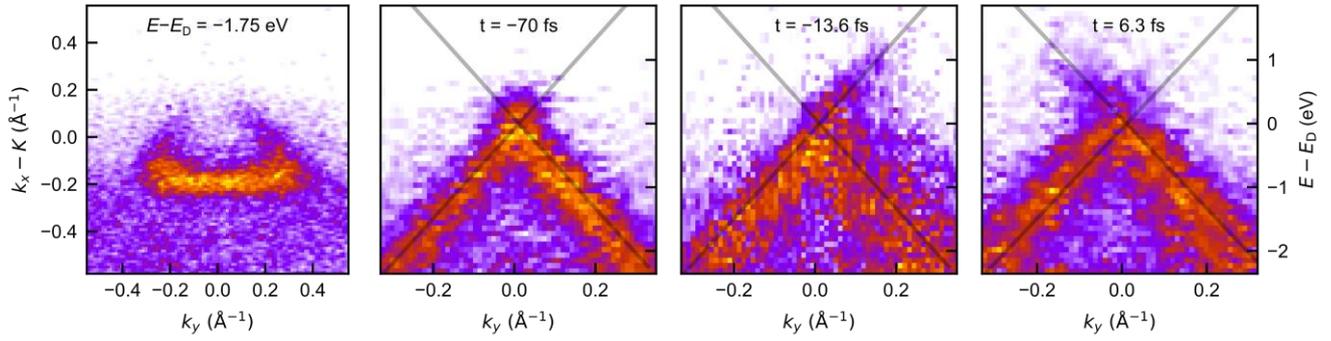

**Extended Data Figure 9 | Raw photoemission data.** Unfiltered version of the band structure data shown in Fig. 2. The first panel shows a zoomed-in momentum map of a cut through the band structure 1.75 eV below the Dirac crossing. The three right panels show the streaking-compensated band structure of graphene at the Dirac point before the arrival of the MIR pulse (t = −70 fs) and for selected delay times during the interaction with the MIR pulse, t = −13.6 fs and t = 6.3 fs.